\documentclass[journal]{IEEEtran}

\usepackage{cite}
\usepackage{graphicx}
\usepackage{amsmath}
\usepackage{amssymb}
\usepackage{amsfonts}
\usepackage{mathtools}
\usepackage[ruled,vlined,linesnumbered]{algorithm2e}
\usepackage{algorithmic} % 如不需要，可移除
\usepackage{url}
\usepackage{float}

\usepackage[table,svgnames,HTML]{xcolor}
\definecolor{green}{HTML}{1A8033}
\definecolor{blue}{HTML}{4DB3E6}
\definecolor{red}{HTML}{E64D4D}

\usepackage{tabularx}
\usepackage{multirow}
\usepackage{makecell}
\usepackage{booktabs}

\usepackage{setspace}
\usepackage[hidelinks]{hyperref}

\begin{document}

\title{Embodied Intelligent Wireless (EIW): Synesthesia of Machines Empowered Wireless Communications}

\author{
Xiang Cheng, Weibo Wen, Haotian Zhang, Boxun Liu, Zonghui Yang, Jianan Zhang, Xuesong Cai

%  W. Wen, H. Zhang, B. Liu, Z. Yang, J. Zhang and X. Cai
\thanks{X. Cheng, W. Wen, H. Zhang, B. Liu, Z. Yang, J. Zhang and X. Cai are with the State Key Laboratory of Photonics and Communications, School of Electronics, Peking University, Beijing 100871, China (email: \url{xiangcheng@pku.edu.cn}; \url{weber@stu.pku.edu.cn}; \url{haotianzhang@stu.pku.edu.cn}; \url{boxunliu@stu.pku.edu.cn}; \url{yzh22@stu.pku.edu.cn}; \url{zhangjianan@pku.edu.cn}; \url{xuesong.cai@pku.edu.cn}).}
}

\markboth{Journal of \LaTeX\ Class Files,~Vol.~X, No.~X, November~2025}%
{Journal of \LaTeX\ Class Files,~Vol.~X, No.~X, November~2025}

\maketitle

\begin{abstract}

The evolution toward the sixth-generation (6G) and beyond mobile communication systems is marked by a fundamental shift from merely connecting devices to enabling pervasive and embodied intelligence. While recent advances in artificial intelligence (AI)-native wireless communication designs have achieved remarkable progress, the prevailing paradigm remains limited to static, modular AI substitutions. This approach fails to meet the core requirements of future wireless networks: the ability to continuously perceive, adapt to, and interact with the dynamic wireless environment. 
To bridge this gap, this paper introduces Embodied Intelligent Wireless (EIW), a novel communication paradigm inspired by embodied intelligence, which redefines the communication node as an active, environment-aware, and evolving entity. EIW is built around an observation–decision–action paradigm, comprising: multi-dimensional observations for comprehensive awareness of the environment and system states, a unified decision module for orchestrating multiple wireless agents in an interpretable manner, and actions where wireless agents exert practical effects on both the environment and communication systems. Furthermore, two enabling technologies, wireless world models as well as self-update and self-evolution mechanisms, are introduced to support training efficiency improvement, counterfactual evaluation, and better adaptation. Unlike existing communication systems, EIW envisions future communication systems not as passive data pipelines, but as intelligent entities that continuously interact with and co-evolve alongside their environments. Finally, through 
simulations, we showcase the advantages of the proposed EIW paradigm and its enabling
techniques in shaping the design of wireless communication nodes.

\end{abstract}

\section{Introduction}
\IEEEPARstart{W}{h}ile the commercial deployment of the fifth-generation (5G) mobile communication technology is still underway, research toward the sixth generation (6G) has gained full momentum, targeting the extreme demands for wireless connectivity in the 2030s and beyond \cite{6G}. Emerging wireless network visions, such as ubiquitous intelligence, digital twins, and holographic interaction, impose unprecedented requirements on key performance indicators, including peak rate, latency, and reliability. They also require the network to concurrently sustain diverse services (e.g., large-scale sensor data uplink and mission-critical command delivery). These services are heterogeneous by nature, and their quality-of-service (QoS) demands vary over time. Consequently, communication systems that rely on conventional designs based on fixed modeling assumptions struggle to perceive and adapt in real time to such highly heterogeneous and dynamic QoS requirements. Moreover, in complex environments (e.g., dense urban areas and smart factories), time-varying wireless channels, severe interference, and scarce spectrum resources further exacerbate the difficulty of guaranteeing high QoS, raising the bar for robust, efficient, and adaptive system design under stringent constraints. To address these fundamental challenges, \textbf{\textit{Synesthesia of Machines (SoM)}} \cite{SoM} has been proposed as a promising enabling technology for future wireless communication systems. SoM aims to intelligently integrate communications and multi-modal sensing through AI-native processing, offering a pathway to more adaptive and efficient network operation. Two core features of SoM are highlighted below:

% These challenges expose fundamental limitations of the prevailing technological paradigm and highlight the need for a conceptual shift: a \textit{next-generation wireless communication paradigm} capable of proactively sensing physical environment, wireless channel state and system service state, making intelligent decisions, and shaping the wireless environment and systems in real time.

\textbf{\textit{Integrated AI and Communications}}: Artificial intelligence (AI) serves as the core tool for SoM. AI technologies are progressively being embedded into wireless networks. This trend has gained recognition from international standards bodies~\cite{ai}. For instance, 3GPP has included AI for the air interface (AI-native air interface), specifically AI-based beam management and channel state information (CSI) feedback enhancement, as study items in Release-18, signifying that the deep integration of AI and wireless communication is an irreversible future direction. However, conventional deep learning models exhibit limited transferability beyond their training distributions. Trained in a supervised manner on specific tasks and datasets, they tend to overfit to dataset biases, degrade sharply under domain shifts (e.g., new channels, mobility patterns, hardware impairments), and require costly re-labeling and retraining for each new scenario. Against this backdrop, the fundamental AI methodology for SoM is shifting from ``one-task-one-model” to \emph{Wireless Foundation Models}~\cite{tnsewifo}. Compared with task-specific models, wireless foundation models increase capacity and leverage pretraining to offer transferability for wireless tasks. Furthermore, a single foundation model can serve multiple wireless tasks, significantly reducing the number of required models. They also adapt efficiently to domain shifts with limited data and fine-tuning, thereby reducing data collection and adaptation costs.

\textbf{\textit{Integrated Multi-Modal Sensing and Communications}}: A central aspect of SoM is the exploitation of multi-modal sensing information. Most existing wireless communication algorithms remain radio-frequency (RF)-only, relying mainly on pilot transmission to obtain channel state. However, such RF-only observation provides only a limited view of the environment, failing to capture richer information such as 3D structure of the scene, user and device spatial distributions, mobility patterns, blockage likelihoods, material properties, or even upstream indicators like traffic demand forecasts. Alongside the continuous rise in intelligence levels across industries, multi-modal sensing is becoming increasingly pervasive. In fields like autonomous driving, intelligent logistics, and smart homes, agents are commonly equipped with heterogeneous sensors such as cameras, light detection and ranging (LiDAR), Global Positioning System (GPS), and millimeter-wave (mmWave) radar to build a multi-dimensional understanding of the physical world. This presents a new opportunity for wireless communication systems: communication nodes can potentially transcend their conventional single-perspective view based solely on RF signals~\cite{ISAC}. They can open other eyes by fusing multi-modal sensing data from other frequency bands of electromagnetic waves to observe the physical world from different dimensions like geometry, material, and motion. SoM systematically guides this deep integration, exploring the technical roadmap and underlying principles.

\textbf{\textit{The Gap Toward Truly Intelligent Wireless Systems}}: 
% Despite the advances outlined above, a significant gap remains between current capabilities and the vision of a fully intelligent, adaptive wireless system. 
Despite the promise of SoM as a potential enabler, a significant gap remains between its current capabilities and the vision of a fully intelligent, adaptive wireless system~\cite{agent}.
Whether based on traditional model-driven algorithms or contemporary AI-based approaches, today's wireless communication systems operate under a fundamentally static paradigm. Even in state-of-the-art implementations, core operational parameters and behavioral logic are typically fixed during pre-deployment optimization and remain largely unchanged during runtime. Consequently, they lack the inherent capability to dynamically optimize or fundamentally alter their strategies in response to real-time changes in the radio environment, user distribution, or heterogeneous service demands. While foundation models represent a leap forward in generalization, their performance is ultimately bounded by the scope and diversity of their pretraining data. Superior generalization relative to task-specific models does not guarantee satisfactory performance in all unforeseen scenarios. Therefore, even the most advanced models today can be viewed as sophisticated yet ultimately static mappings. Furthermore, the integration of such AI models introduces new operational complexities, including model updating, mode switching, and runtime scheduling, underscoring the need for an efficient intelligent decision core capable of unified model orchestration. To bridge this gap, the next leap in wireless intelligence has to transition from static models toward systems that can autonomously refine their algorithms and policies based on live environmental observations to achieve enduring efficiency and robustness in dynamic environments~\cite{embodied}. 

In this paper, we propose the \textbf{\textit{Embodied Intelligent Wireless (EIW) }}as the next-generation wireless communication paradigm inspired by embodied intelligence. We first introduce the motivation and processes of the EIW paradigm. Then, we present several key mechanisms, including multi-dimensional observation, unified decision for model orchestration, and wireless foundation model-empowered action, to realize EIW. Moreover, two enabling techniques are discussed and a case study is conducted to validate the effectiveness of the proposed EIW paradigm. Finally, the open issues and future research directions are outlined.

% Within the architecture of EIW, communication systems are no longer just passive data transmission pipelines, but embodied entities that can continuously interact with the environment, learn and evolve, achieving a paradigm shift from passive response to active environment and system shaping.  

\section{EIW: A New Paradigm for Wireless Communications}

Embodied intelligence is an advanced AI paradigm that enables intelligent agents to perform physical tasks through real-time interaction with the environment. Unlike traditional AI, which primarily learns from offline datasets and operates within predefined input–output mappings, embodied intelligence emphasizes online perception, active decision-making, and physical action in dynamic and open worlds.
Typical processes of embodied intelligence, for example, in robots and autonomous vehicles, follow an observation–decision–action paradigm. During observation, the agent acquires multi-modal environmental data via sensors such as cameras, LiDAR, inertial measurement units (IMU), and microphones. The decision module involves interpreting the observation, evaluating possible strategies, and generating executable plans. Finally, in the action process, the agent influences the physical world through actuators such as robotic arms and mobile platforms.

% \subsection{Embodied Intelligent Wireless}

\begin{figure*}[!t]
    \centering
    \includegraphics[width=\linewidth]{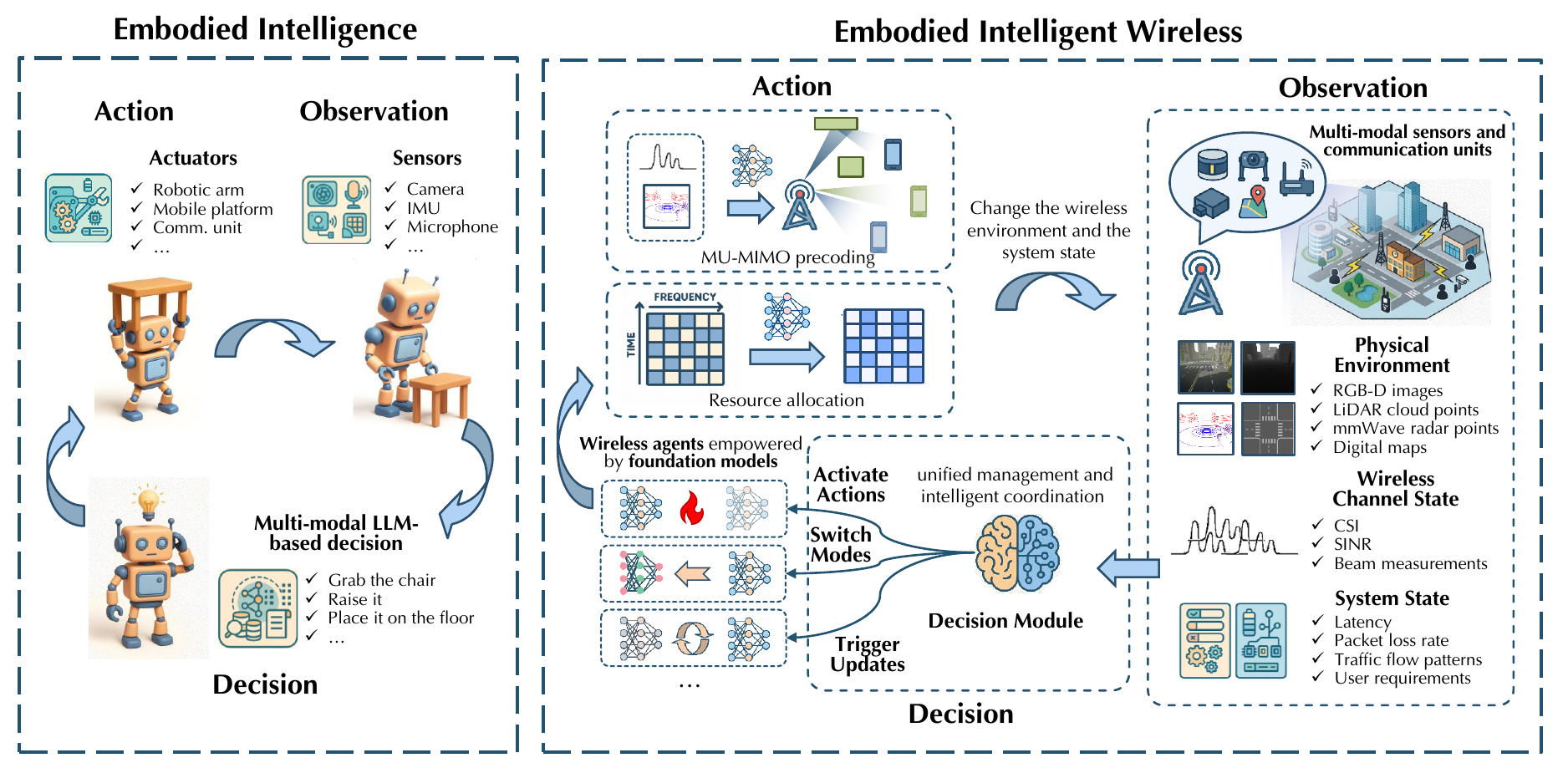}
    \caption{Illustration of EIW paradigm.}
    \label{fig:EIW-arch}
\end{figure*}

Inspired by embodied intelligence, we propose EIW, where wireless communication nodes are designed as embodied intelligent entities. Specifically, the embodied entities learn a joint representation of the physical and wireless environment from multi-dimensional observations, and then make decisions and take actions that optimize communication utility, including throughput, latency, reliability, and energy efficiency. EIW consists of three processes: (i) \emph{observation}, (ii) \emph{decision}, and (iii) \emph{action}, as illustrated in Fig.~\ref{fig:EIW-arch}.

% 怎么感觉写的有点像intro

%\subsubsection{Architecture}

% Guided by SoM, EIW aims to conduct comprehensive observations of the physical and electromagnetic environments, as well as the state of communication systems. These observations are multi-dimensional, encompassing the \emph{physical environment, wireless channel state, and system state}.
% Physical environment information is obtained via on-device sensors—e.g., mmWave radar for RF scene sensing and cameras or LiDAR for non-RF perception—to capture geometry, motion, and occlusions around the transceivers. Some external sources provide supplementary priors, such as digital maps and bird’s-eye-view (BEV) feature maps.
% Wireless channel state comprises direct observations of the radio link, such as pilot signals, beam-sweeping feedback, received power, CSI, and signal-to-interference-plus-noise ratio (SINR).
% System state primarily includes protocol- and traffic-related indicators, including physical resource block (PRB) utilization and grants, scheduling and access status, throughput, and latency.
% These diverse information sources enable EIW to form a more comprehensive understanding of the wireless environment and system than traditional communication systems.

 \emph{1) Observation}: Guided by SoM, EIW aims to conduct comprehensive observations of the physical and wireless environments, as well as the state of communication systems. These observations are multi-dimensional, encompassing the \emph{physical environment, wireless channel state, and system state}. These diverse information sources enable EIW to form a more comprehensive understanding of the wireless environment and system than traditional communication systems.

% AI-native communication systems are facing a new set of requirements that differ fundamentally from traditional protocol control. These AI models introduce new problems such as determining when to update or switch models, identifying the appropriate operating mode, and diagnosing which model constitutes the system bottleneck. The decision module in EIW is designed to address these challenges by enabling unified management and intelligent coordination of AI models across the communication protocol stack. 
% It has knowledge of the dynamic characteristics, interrelationships, and impact on overall performance of various modules across the communication protocol stack. The decision module can also analyze real-time multi-dimensional observations to construct a unified representation of the global system state. Leveraging this deep understanding of the global state, the decision module outputs high-level directives that serve to achieve unified management and intelligent coordination of AI models, guiding wireless agents in the physical (PHY) and media access control (MAC) layers to take appropriate actions.

\emph{2) Decision:} The decision process transforms the observation representation into high-level directives that orchestrate the behavior of AI models across the communication stack. Unlike conventional systems, where PHY and MAC procedures are designed and configured largely in isolation under a fixed protocol, EIW employs a unified decision module that reasons about the global impact of different configurations, model mode choices, and update strategies. By learning the dynamic relationships among modules and evaluating trade-offs such as throughput versus latency or reliability versus computational cost, this module issues abstract control signals that flexibly steer how multiple wireless agents should operate.

\emph{3) Action:}  EIW makes practical effects on both the environment and the communication systems via actions. Actions are taken primarily by the wireless agents realized by wireless foundation models, while the processes are triggered by directives issued from the decision module. For example, a model may be invoked once for inference to update beamforming vectors or to reselect the modulation and coding scheme (MCS), thereby altering the wireless environment and the system state in ways that benefit transmission. Another form of action is performed by computation-oriented models that operate continuously to sustain link stability—for instance, through channel estimation and signal detection. Furthermore, models may undergo scheduled update procedures, such as fine-tuning or meta-adaptation, to preserve adaptability under dynamic wireless conditions.

\begin{figure*}[!t]
    \centering
    \includegraphics[width=1.0\linewidth]{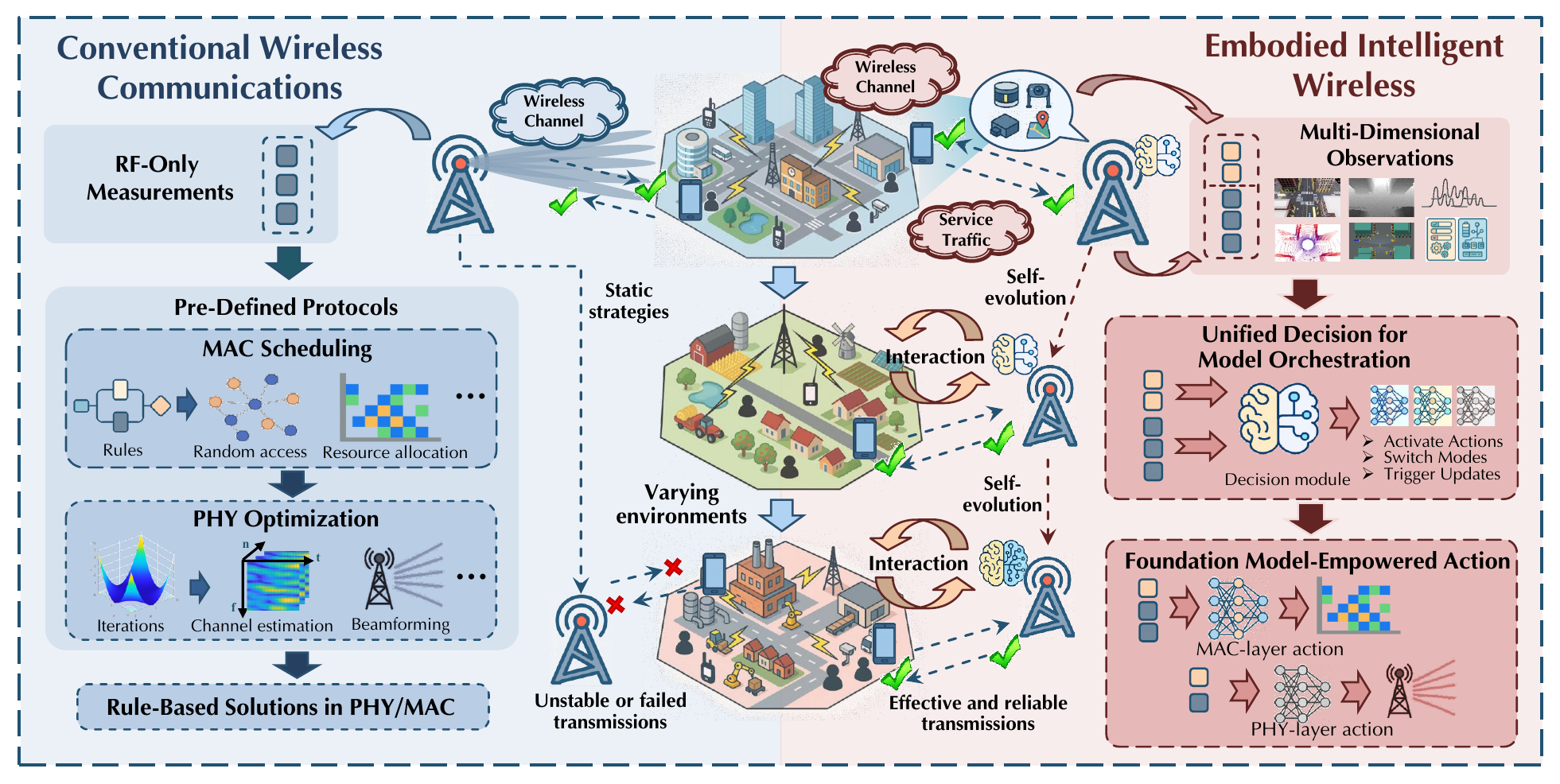}
    \caption{Comparisons between EIW and traditional wireless communications.}
    \label{fig:EIW-comp}
\end{figure*}

In summary, EIW represents an emerging paradigm that stands apart from traditional wireless communication systems in its design. The key design distinctions between EIW and conventional systems are shown in Fig.~\ref{fig:EIW-comp}.

\section{Mechanisms and Enabling Techniques for EIW}

In this section, we elaborate on the mechanisms of EIW and introduce several key enabling technologies required for its operation.

\subsection{Multi-Dimensional Observation}
% Unlike traditional wireless systems that primarily rely on signal measurements to infer the radio environment, the hallmark of EIW is its multi-modal observation capability. EIW can acquire rich, fine-grained observations across three key dimensions: (i) \emph{physical environment}, (ii) \emph{wireless channel state}, and (iii) \emph{system state}. This multi-dimensional observation enables EIW to construct a comprehensive environmental awareness map that goes beyond the traditional ``channel-centric” view, encompassing the physical space, the electromagnetic space, and the user state space. As a result, it provides more reliable and context-rich information to support subsequent decision and action processes~\cite{multi-modal}.

Unlike traditional wireless systems that infer the radio environment solely from signal measurements, EIW leverages multi-modal observations across three key dimensions: (i) the \emph{physical environment}, (ii) the \emph{wireless channel state}, and (iii) the \emph{system state}. By integrating these multi-dimensional observations, EIW constructs an environmental awareness map that extends beyond the conventional channel-centric perspective and provides context-rich information for subsequent decision and action processes~\cite{multi-modal}.

\subsubsection{Physical Environment Observation}
Physical environment observation refers to the direct acquisition of geometric and semantic information about the physical environment through various sensors (e.g., LiDAR, cameras, mmWave radar, and digital map) deployed on the communication nodes. Specific data includes high-precision 3D point clouds, texture- and semantic-rich 2D images, and information on target distance, velocity, and angle \cite{dataset}. Under the guidance of SoM, physical environment observation is no longer isolated information but is instead processed in conjunction with wireless data to reveal the complex mapping between physical and electromagnetic spaces. This process provides powerful geometric and semantic priors for understanding the wireless propagation environment. For example, by identifying obstacles, reflective surfaces, and scatterers, EIW can take advantage of the learned mapping relationship to predict potential signal blockages or reflection paths, thereby assisting in proactive beam management and handover decisions \cite{MMFF}. In this process, physical environment observation is transformed by EIW into communication performance gains.

\subsubsection{Wireless Channel State Observation}
Wireless channel state observation refers to real-time channel information over the communication link, such as demodulation reference signals and sounding reference signals. This observation represents the most direct and fundamental way to monitor the wireless channel. Key parameters include CSI, SINR, and beam measurement results (e.g., reference signal received power). Wireless channel state observation serves as the direct foundation for physical layer and link layer optimization. It not only acts as the underlying input for decision but also provides a basis for integrating physical environment observations. Based on wireless channel state observations, EIW can accurately evaluate link quality, offering core data support for critical tasks such as adaptive modulation and coding (AMC), precise beamforming, and interference coordination.

\subsubsection{System State Observation}
System state observation focuses on monitoring the system status and service demands. It gathers information such as user activity status and a range of QoS-related indicators, including latency, reliability, and service demand levels. System state observation concretizes users' communication demands and directly impacts resource scheduling and optimization strategies at the PHY and MAC layers, including key operations such as AMC, user scheduling. By understanding the real-time requirements of different services (e.g., eMBB, uRLLC, mMTC), EIW enables QoS-aware intelligent decisions and actions.

\subsection{Unified Decision for Model Orchestration}

The decision module serves as the central brain of a communication node. Its role is to analyze the live state of the communication system and to output coordination and optimization strategies for all models. These strategies are issued as high-level directives that direct AI models at the PHY and MAC layers to act in a coordinated way. The directives can be grouped but not limited into three types: activate actions, switch models or modes, and trigger updates. First, activate actions are one-shot calls made on demand. For example, when many new users arrive within a short time, the decision module immediately starts a time–frequency resource-allocation step to reassign PRBs and to refresh scheduling and admission thresholds, so as to avoid congestion and protect the QoS of key traffic. Likewise, when channel quality drops (e.g., a lower SINR or stronger interference), a link-adaptation model is triggered to run inference. Second, the decision module can switch models or change their run modes as needed. It may adjust the input modalities used by some models (for instance, it may down-weight visual cues under occlusion in a task that uses multi-modal sensing), or change the mode to match the scenario; for example, after detecting open terrain with few multi-path components, it can increase the length of the predicted sequence of a channel prediction model, or pick a lightweight or distilled version to save computing resources. Third, the decision module can issue update directives that start a preset model-update process. If a sharp drop in system performance is traced to one model, it indicates that the model no longer fits the current scenario and has become a bottleneck. In response, the module collects the training samples required by this model and performs a targeted few-shot update or meta-adaptation to help the model adapt to the present scenario.

\subsection{Wireless Foundation Model Empowered Action}

The actions are executed by multiple \textit{wireless agents}, each responsible for one or more PHY- or MAC-layer functions that implement the basic capabilities of the wireless communication system. In principle, each wireless agent can be realized as a parameterized model, a task-specific AI model, or a wireless foundation model \cite{tnsewifo,wifo}. Wireless foundation models are adopted as wireless agents in EIW for three reasons. First, wireless foundation models provide sufficient modeling capacity to exploit sensing information and operate in dynamic environments. Second, because of their large capacity, wireless foundation models exhibit strong generalization across diverse environments, thereby reducing the cost of updates and fine-tuning. Third, conventional modules or task-specific AI models would require many distinct wireless agents, increasing management and decision-making overhead, whereas a wireless foundation model allows a single wireless agent to support multiple PHY and MAC tasks.

Actions are grouped into three types: single-shot inference, sustained operation, and in-time update. Single-shot inference denotes actions that execute a specific task according to directives from the decision module, such as  switching MCS or allocating resources to new users. These actions help the system rapidly adapt to changes in the wireless environment. For example, when the communication node observes SINR degradation, the wireless foundation model infers the modulation and coding scheme, and the result is translated into control commands for the physical-layer transceivers to realize real-time MCS adaptation and restore link quality. Sustained operation covers PHY and MAC functions that run continuously, including channel estimation, equalization, signal detection, channel coding and decoding, and CSI feedback. For instance, a foundation-model-based channel estimator can be executed once per time slot during downlink transmission to maintain up-to-date CSI for data transmission~\cite{wifo}. In-time update denotes adaptation of the action models themselves when the environment changes. Although a wireless foundation model has broad generalization, it can still encounter out-of-distribution conditions; meta-learning and fine-tuning are thus used for fast adaptation of the wireless agents, maintaining performance across communication contexts. Through these three types of actions, wireless agents respond to directives from the decision module, preserve key communication functions, and update to meet dynamic wireless conditions.

\begin{figure*}[ht]
    \centering
    \includegraphics[width=0.9\linewidth]{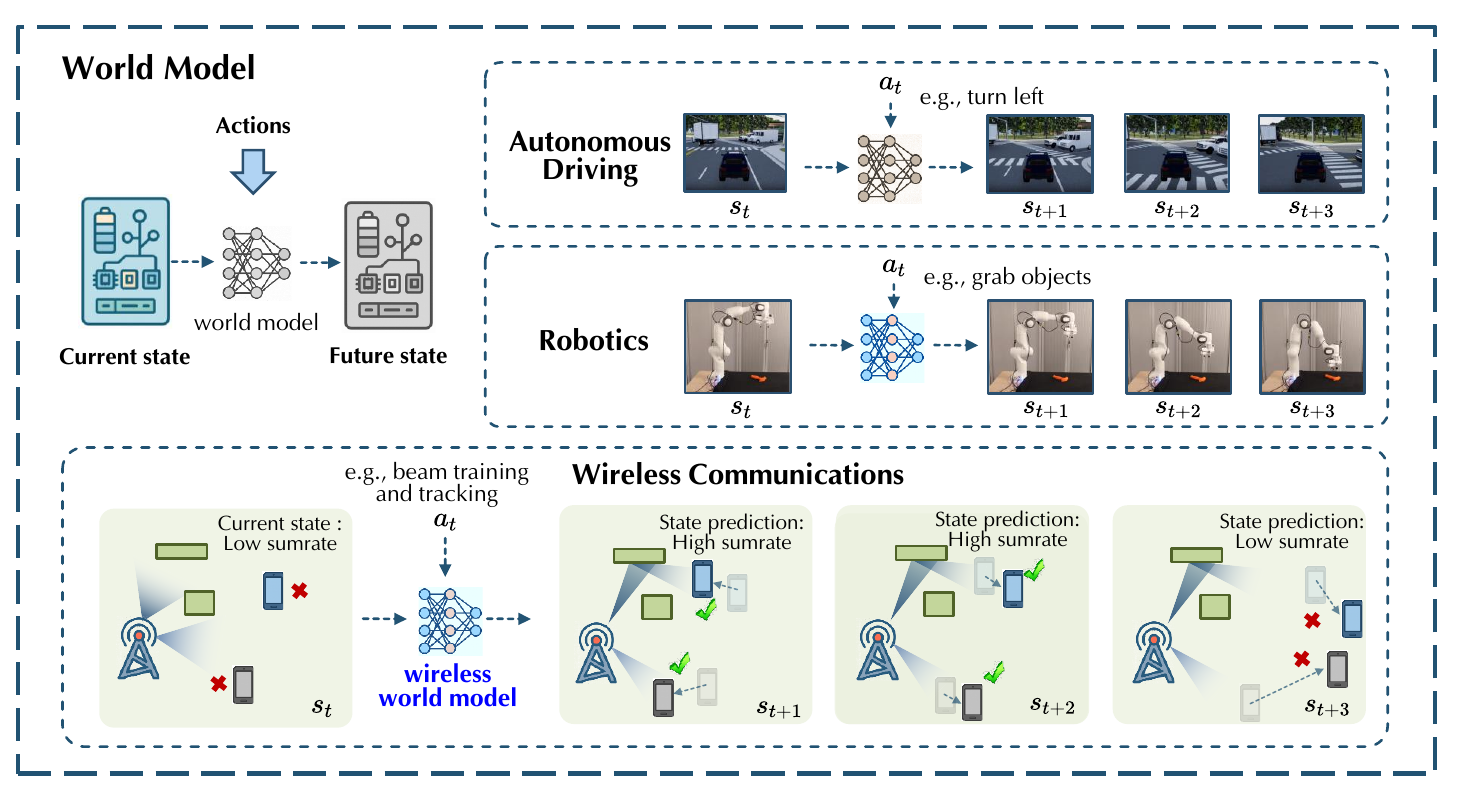}
    \caption{Illustration of wireless world model.}
    \label{fig:WWM}
\end{figure*}

\subsection{Enabling Techniques}

\subsubsection{Wireless World Model}
World models are a class of models that enable an agent to learn and internally represent the structure and dynamics of its environment, thereby characterizing how states evolve under actions~\cite{worldmodels}. Unlike world models in robotics or autonomous driving that describe interactions with the physical world, \textit{wireless world models} capture and predict the dynamics of the wireless environment in which wireless agents operate. The wireless world model can be characterized by the following features:
\begin{itemize}
\item \textbf{Environment States:} The wireless world model focuses on the stochastic and temporal evolution of wireless environment and system states, rather than geometric or kinematic properties of the physical world.
\item \textbf{Task Objectives:} The wireless world model focuses on facilitating communication tasks driven by performance metrics such as spectral efficiency, throughput, reliability, and resource utilization, thereby guiding intelligent actions in dynamic wireless environments.
\end{itemize}

Functionally, the wireless world model can serve as either an evaluator or a simulator. As an evaluator, it predicts the future system state based on the current channel and resource-allocation status, together with the actions generated by a wireless foundation model. Through this look-ahead prediction, the system can perform counterfactual evaluation and filter unsafe actions, thus preventing potential degradation in real wireless operations. For example, if the wireless world model forecasts that a beam-switching action would reduce a user’s SINR, such an action can be rejected or replaced. As a simulator, the wireless world model substitutes for the real environment to provide supervisory signals in reinforcement learning (RL), enabling the training of deep learning (DL) models without costly or risky interactions with real-time wireless systems~\cite{worldmodel}. For instance, a link-adaptive action model can be trained on data generated by a wireless world model adapted to the current scenario, rather than exploring in the live network. In both roles, the model can be continually updated with real-time measurements to reduce the sim-to-real gap and enhance sample efficiency.

% The wireless world model can serve as a simulator to assist in the training of other models as well. The underlying principle is that the world model substitutes for the real environment to provide supervisory signals in reinforcement learning (RL). This allows deep learning (DL) models to avoid direct interaction with real-time running systems in each iteration, where obtaining sample trajectory can significantly reduce high cost and potentially dangerous environmental interactions~\cite{worldmodel}. For example, a link adaptive action model based on reinforcement learning can be trained using samples generated from a wireless world model that has adapted to the current scenario, without the need to explore in real wireless systems. In both evaluator and simulator roles, the world model can be continually updated with real-time measurements to shrink the sim-to-real gap and improve sample efficiency.

\subsubsection{Self-Update and Self-Evolution} 

The EIW imposes more stringent requirements on the capability of wireless foundation models to autonomously adapt to their operating environments. From a short-term self-update perspective, the EIW must independently assess new wireless conditions and achieve rapid adaptation. In the wireless domain, such fast adaptation can be broadly categorized into two classes. The first class involves parameter-efficient fine-tuning techniques, such as adapter-based fine-tuning and low-rank adaptation (LoRA). The second class corresponds to meta-learning~\cite{meta}, with the most representative example being Model-Agnostic Meta-Learning (MAML). The core principle of meta-learning for fast adaptation is to incorporate the model’s few-shot adaptability to new environments as an explicit supervisory signal in the optimization processes.

From a long-term self-evolution perspective, the EIW must possess the capability to continually acquire new wireless knowledge without suffering from catastrophic forgetting of previously learned information~\cite{continual}. This necessitates the adoption of continual learning techniques. Among the prevailing strategies, regularization-based approaches impose constraints on parameter updates or output distributions to preserve previously acquired knowledge, mitigating catastrophic forgetting.
Replay-based methods maintain performance on earlier tasks by either storing a subset of past samples (experience replay) or reconstructing them through generative models (generative replay).
Architecture-based approaches allocate task-specific parameters via modular expansion or parameter isolation mechanisms, enabling scalable and incremental knowledge acquisition. These approaches empower communication nodes to incrementally acquire, update, accumulate, and utilize knowledge throughout their operational lifespan.

\section{Case Study}

\begin{figure*}[!ht]
    \centering
    \includegraphics[width=0.95\linewidth]{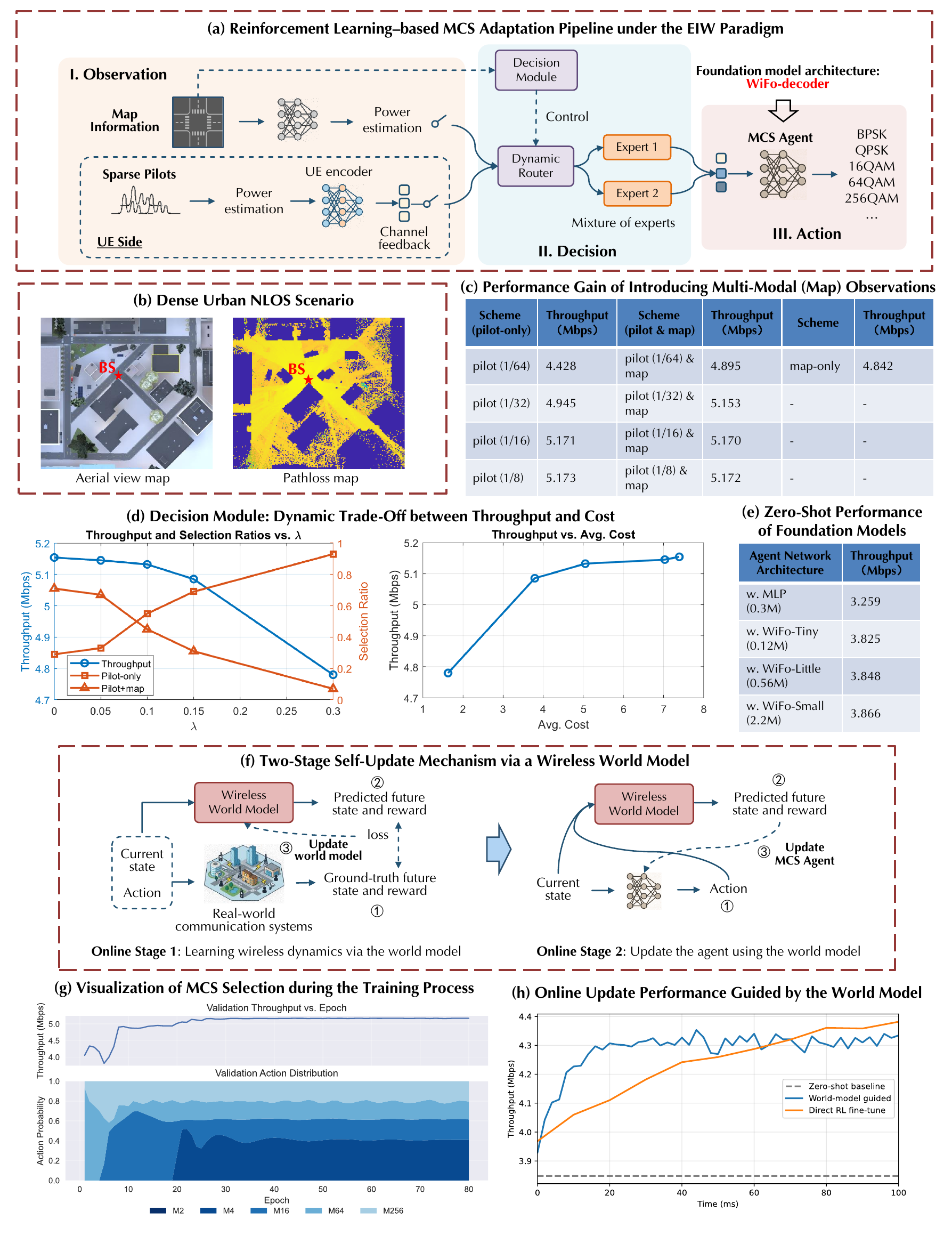}
    \caption{(a) Illustration of the reinforcement-learning–based MCS adaptation pipeline under EIW paradigm; (b) the dense urban NLOS environment with aerial and pathloss maps; (c) comparison of throughput for pilot-only, map-only and pilot+map schemes; (d) the decision-module trade-off between throughput and computational cost under different expert-selection ratios; (e) zero-shot throughput results of small model and foundation models of various sizes; (f) the two-stage self-update mechanism via a wireless world model; (g) visualization of the distribution of MCS selection and throughput during training; (h) throughput over time under different online update schemes.}
    \label{fig:Case}
\end{figure*}

In this section, we introduce a case study to demonstrate the advantages of the proposed EIW paradigm and its enabling techniques in the design of wireless communication nodes.

\subsection{Reinforcement Learning–Based MCS Adaptation}

We consider a single-base station (BS), single-user downlink multicarrier system. In each transmission time slot, the user receives pilot signals transmitted by the BS on a subset of subcarriers, compresses the coarse SNR estimates into compact features, and feeds these features back to the BS. The BS then selects the modulation order based on the user feedback and applies it to all subcarriers, while the system objective is to maximize the overall transmission rate. As illustrated in Fig.~\ref{fig:Case}(a), we introduce the following design components under the EIW paradigm:

\begin{enumerate}

\item Observation: In addition to the user feedback features that contain SNR estimation information, the BS also introduces map information to enhance the estimation process. The map information comprises $5$ channels: the three RGB channels of an aerial-view map (as shown in Fig.~\ref{fig:Case}(b)), a user-location mask, and a BS-location mask. After being processed by the map encoder, these channels yield the estimated received power from the BS to the user.

\item Decision: A decision module combined with a mixture-of-experts (MoE) mechanism is incorporated to determine the appropriate modality. The lightweight module dynamically routes the input according to the map information. When selecting only pilot-based features, the map-information branch remains inactive, resulting in low computational cost and processing by expert~1. When selecting both pilot and map information, the larger number of parameters in the map encoder provides potentially higher performance but incurs greater computational cost, in which case expert~2 is activated. By adjusting the regularization factor $\lambda$, the decision module governs the trade-off between computational efficiency and model performance.

\item Action: Leveraging the strong reasoning and generalization capabilities of wireless foundation models, the WiFo-decoder~\cite{wifo} is employed as the backbone of the MCS agent (policy model). The agent processes the features corresponding to the modality selected by the expert and outputs the modulation order, which is then applied to all subcarriers in the current time slot.

\end{enumerate}

\subsection{Self-Update via Wireless World Model}

When the wireless communication environment changes, the performance of the policy model that has been pretrained can degrade due to distribution shift. To address this issue, we propose an online optimization framework in which a learned wireless world model is used to continuously adapt the policy model. The world model takes as input the current state and action, where the current state consists of a history of stacked power estimates obtained from map information together with user feedback, and outputs both the next-state transition and the reward associated with the given state–action pair. During pretraining, the policy model and the world model are simultaneously trained. After deployment, whenever changes in the wireless environment indicate that the model needs to be updated, an online update mechanism is activated. As illustrated in Fig.~\ref{fig:Case}(f), this mechanism performs a two-stage iterative optimization. In Stage~1 (learning wireless dynamics via the world model), the world model is continuously calibrated so that its predicted future states and rewards closely match those generated by the real communication system. In Stage~2 (updating the agent using the world-model-predicted reward), the policy is refined based on the rewards provided by the world model rather than directly relying on feedback from the real-world communication system, thereby adapting the agent to the current environment. This design offers two primary benefits: First, it prevents unsafe reinforcement-learning exploration in real communication systems, as potentially harmful or performance-degrading exploratory actions can be evaluated within the simulated world model; second, it substantially improves exploration efficiency, as the agent can perform a large number of simulated roll-outs and policy updates at low cost and with accelerated learning speed, without being constrained by the limited interaction rate and latency requirements of the real communication system.

\subsection{Simulation Results}

We use the SynthSoM dataset~\cite{dataset} to conduct a series of experiments that demonstrate the advantages of designing communication nodes under the proposed EIW paradigm. The simulation results in Fig.~\ref{fig:Case}(c) show the throughput gain obtained by incorporating map information. When the pilot pattern is relatively sparse (with a pilot density of $1/32$ and below), the pilots alone do not provide sufficiently accurate SNR estimates, and in this regime the introduction of map information as side information effectively improves the overall system throughput. Fig.~\ref{fig:Case}(d) further shows that the decision module can dynamically trade off throughput against computational cost. Specifically, we define the cost of selecting the pilot-only branch as $1$ unit, while the cost of jointly using pilots and map information is set to $10$ units based on their computational load. By tuning the hyperparameter $\lambda$, which controls the penalty on computational cost, the decision module automatically adjusts the fraction of time that the higher-cost branch involving map information is activated. In Fig.~\ref{fig:Case}(e), we evaluate the generalization capability of the proposed agent by pretraining the model in a LoS-dominated scenario and then performing zero-shot inference in a NLoS-dominated scenario. Compared with a smaller model (multilayer perceptron), the agent equipped with the WiFo-decoder backbone achieves higher performance.

In addition, the wireless world model exhibits clear sample-efficiency advantages during online self-update. We consider a base station that performs MCS exploration and updates its policy every $10$ ms as part of the online RL procedure. Under the world-model–guided scheme, each model update requires only about $2$ ms on a single NVIDIA RTX 4090 GPU. In Fig.~\ref{fig:Case}(h), compared with a direct RL fine-tuning baseline that explores policies solely through interactions with the real-world communication system, the world-model-guided scheme converges significantly faster and requires far fewer real environment interactions.

\section{Open Issues}
\subsection{Dataset Construction}

The training of the decision module, wireless world models, and wireless foundation models in EIW relies on large amounts of high-quality data. However, existing datasets in the wireless communication domain fail to simultaneously meet the alignment requirements of EIW across three critical dimensions: the physical environment, wireless channel state, and system state. Therefore, it is necessary to develop a new dataset with multi-dimensional alignment to support the training and validation of EIW-related models.

\subsection{Unified Representation of Observations}

The multi-dimensional observational data utilized by EIW exhibits high heterogeneity across modalities.
The core challenge of how EIW leverages multi-dimensional observations lies in transforming heterogeneous observations into consistent knowledge suitable for subsequent steps. This requires abandoning traditional approaches of multi-modal data alignment and feature extraction. Instead, it calls for the investigation of an end-to-end joint learning framework. Such a framework is capable of autonomously discovering the intrinsic relationships across multi-dimensional observations and mapping them into a unified feature space oriented towards the subsequent decision and action layers.

% \subsection{Multi-Task Foundation Models}

% Wireless foundation models are designed to execute specific tasks at the PHY and MAC layers during actions under the EIW paradigm. Given the diversity and abundance of these tasks, an uncoordinated design may lead to model proliferation and fragmentation. To mitigate this, correlated tasks can be integrated and organized into a limited number of functional categories, each supported by a dedicated multi-task foundation model. This strategy not only reduces fragmentation but also promotes representation sharing among related tasks, enhances computational efficiency, and simplifies model management.

\subsection{Realization of Decision Module}

% , which differs from the conventional training of computational or policy units in communication systems that is grounded in explicit mathematical formulations and optimization objectives
The unified decision module orchestrates multiple wireless agents. The core challenge in realizing this model is to design a training framework that enables the decision module to understand the functions of system components, especially AI modules, and to infer from a high level semantic perspective the mapping between observations and high level directives. This necessitates a new training and inference framework that unifies communication priors, system constraints, and multi-modal observations within a common decision space.

\subsection{Design of Wireless World Models}

Current research on world models has been primarily focused on fields such as multimedia, autonomous driving, and robotics. An important open problem lies in how to effectively transfer this world-model-based paradigm to the domain of wireless communications, enabling the model to learn the evolution dynamics of complex and time-varying environments for accurate prediction of wireless channel state and system state. This research direction is expected to enhance the predictive and adaptive capabilities of communication systems while reducing the cost of real-world interactions during online optimization.

% \subsection{Collaboration of EIW Entities}

% % 
% In practical wireless communication networks, it is significant to introduce interaction and cooperation among multiple communication nodes. The collaboration of these communication nodes can integrate richer information sources from different communication entities, enabling joint decision-making and resource optimization at the global level. Such a cooperative mechanism avoids the local optima that may arise from isolated design and facilitates the achievement of globally optimal communication performance and system-level coordination.

\section{Conclusion}
This paper introduces EIW, a new wireless communication paradigm that transforms communication entities into embodied agents. The mechanisms of EIW integrate multi-dimensional observation, unified decision, and model-driven action to enable adaptive and coordinated operation across diverse demands and environment. Future communication nodes in EIW are no longer passive data pipelines, but intelligent entities that continuously interact with and co-evolve alongside their environments. We envision EIW as the paradigm for next-generation wireless communications and encourage broader exploration of intelligent communication systems built upon it.

%===================== 参考文献都写在本 .tex 文件中 =====================%
% 使用 thebibliography 环境，不需要 .bib/.bst。
% 数字宽度 {00} 可按条目数量增减（两位数字足够写 1–99 条）

\end{document}